\documentclass[12pt]{iopart}
\usepackage{graphics}
\usepackage{hyperref}
\usepackage{xfrac}
\usepackage{graphicx} 
\usepackage[font=small,labelfont=bf]{caption} 
\usepackage[subrefformat=parens]{subcaption}
\usepackage{amssymb}
\usepackage{multirow}
\usepackage[LGRgreek]{mathastext}
\begin{document}

\title[]{Measurement of $^{144}$Sm(p,$\gamma$) cross-section at Gamow energies}

\author{Tanmoy~Bar$^{1,2}$, Dipali~Basak$^{1,2}$, Lalit~Kumar~Sahoo$^{1,2}$, Sukhendu~Saha$^{1,2}$, Jagannath~Datta$^3$, Sandipan~Dasgupta$^3$, Chinmay~Basu$^{1,2}$}

\address{$^1$Nuclear Physics Division, Saha Institute of Nuclear Physics, 1/AF Bidhannagar, Saltlake City, Kolkata~-~700 064, India}
\address{$^2$Homi Bhabha National Institute, BARC Training School Complex, Anushaktinagar, Mumbai~-~400 094, India}
\address{$^3$Analytical Chemistry Division, Bhabha Atomic Research Centre; Variable Energy Cyclotron Centre, 1/AF Bidhannagar, Saltlake City, Kolkata~-~700 064, India}

\ead{tanmoy.bar@saha.ac.in}
\vspace{10pt}
\begin{indented}
\item[]September 2023
\end{indented}

\begin{abstract}
The cross-section measurement of $^{144}$Sm(p, $\gamma$)$^{145}$Eu (T$_{1/2}=$5.93(4) days) reaction has been performed at proton energies around 2.6, 3.1, 3.7, 4.1, 4.2, 4.7, 5.1, 5.5, 5.9, 6.4, 6.8 MeV using the activation technique. These energies correspond to the Gamow window for 3, 4 GK and a partial portion for 2 GK. $^{144}$Sm has been chosen for the present study because of its significantly higher abundance compared to the other neighboring $p$-nuclides (Z $>$ 50). The astrophysical $S-$factor of this reaction has been measured for the first time at E$^{cm}_p=$ 2.57$\pm$0.13 MeV, $S-$ factor$=$2.542$\pm$1.152($\times$10$^{10}$) MeV-b. Cross-section data were compared with the previously measured experimental data from literature and the theoretical predictions obtained using Hauser-Feshbach statistical model codes TALYS 1.96 and NON-SMOKER. A satisfactory agreement between experimental data and theoretical results was observed. Molecular deposition technique was used to prepare the $^{144}$Sm targets having thickness between 100$-$350 $\mu$g/cm$^2$ on Al backing. Obtained results were utilised to predict the reaction rates for $^{144}$Sm (p, $\gamma$) and $^{145}$Eu ($\gamma$, p) reactions using TALYS 1.96 and the reciprocity theorem.
\end{abstract}

%
%
%
%
%

\section{Introduction}
\begin{figure}
  \centering
    \includegraphics[clip, trim=0.0cm 0.0cm 0.0cm 0.0cm,width=0.6\linewidth]{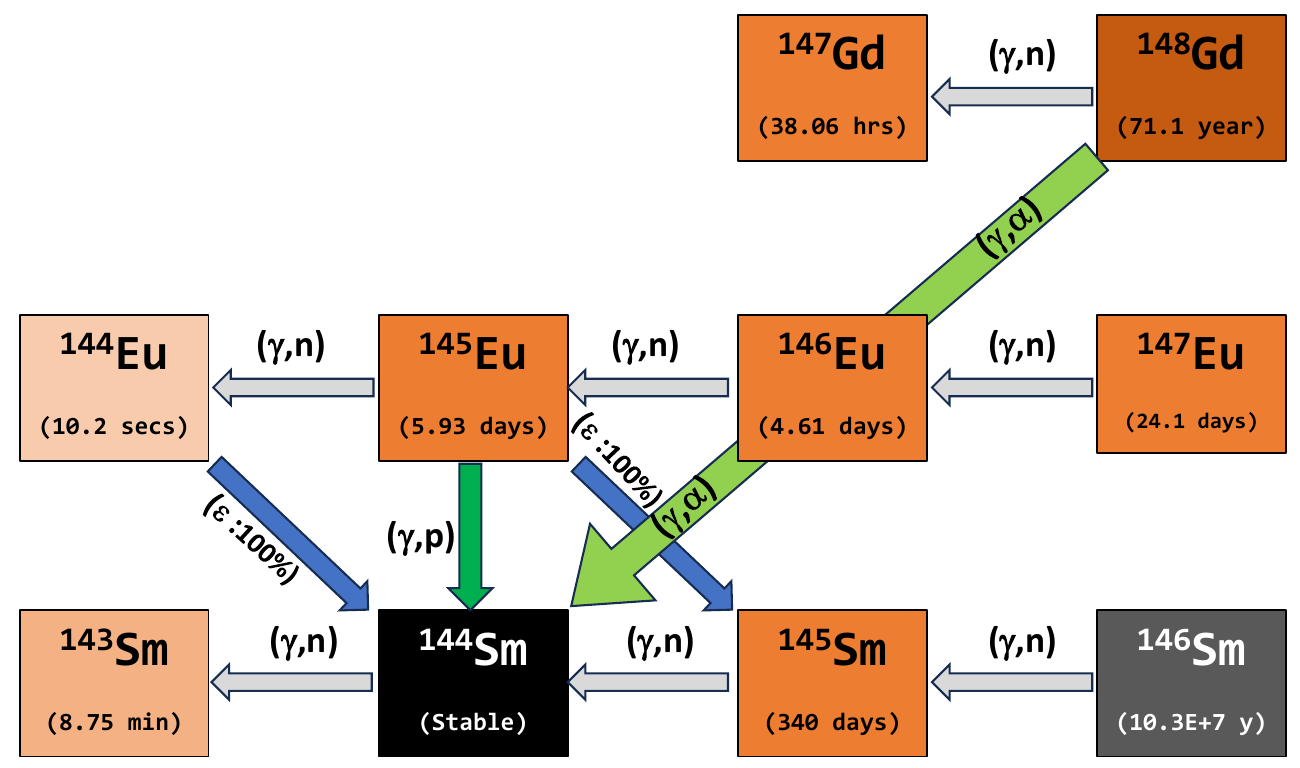}
    \caption{A simplified production pathway of $^{144}$Sm nuclei involving photo-disintegration.} 
    \label{pathway} 
  \end{figure}

One of the primary objectives of experimental nuclear astrophysics is to determine the rates of nuclear reactions that occur in stars under various astrophysical conditions. The reaction rates are determined from the cross sections, which need to be measured at energies as close as possible to the astrophysically relevant ones (near the Gamow window). In many cases, the final nucleus of an astrophysical reaction is radioactive, which allows the cross-section to be determined from offline measurement of the produced isotopes (activation method)~\cite{gyurky2019activation}. Beyond Fe, there is a class of 35 neutron deficit nuclides, between $^{74}$Se and $^{196}$Hg, called p$-$nuclei~\cite{arnould2003p,woosley1978p}. They are bypassed by the $s$ and $r$ neutron capture processes and are typically 10$-$1000 times less abundant than s and/or r isotopes in the solar system. There is a typical abundance of 1\% for lighter nuclei with 34 $\leq$ Z $\leq$ 50 and 0.01$-$0.3\% for medium and heavier nuclei with atomic numbers $>$50. The $p$-nuclei are produced mainly through the $\gamma-$process where either ($\gamma$, n), ($\gamma$, p) or ($\gamma$, $\alpha$) reactions produced it. A high temperature is essential for the $\gamma-$process to take place so as to have a high density of photons. Generally, the abundance of $p$-nuclei decreases with an increase in atomic number, but for neutron magic $p$-nuclei $^{92}$Mo and $^{144}$Sm, it is 14.52\% and 3.08\%, respectively. More detailed and precise information on the reaction cross-section of these nuclei in the astrophysical energy region is extremely important. $^{144}$Sm is an important $p$-nuclei, that shows a much higher solar abundance in comparison to other $p$-nuclei (it is second highest in abundance after $^{92}$Mo) due to its magic neutron shell. A formation path of $^{144}$Sm is shown in figure~\ref{pathway} where it is seen to be produced by ($\gamma$, p) process. As high intensity $\gamma-$beams are not easily available around the world, the inverse (p, $\gamma$) reaction is easier to measure and use the principle of detailed balance to obtain the ($\gamma$, p) reaction rate.  In this paper, the $^{144}$Sm(p,$\gamma$) reaction cross-section is reported in the astrophysically important energy region (T$_9=$2 and 3). Previously N. Kinoshita \textit{et al.}~\cite{kinoshita2016proton} has performed this experiment in the 2.8$-$7.5 MeV energy regime using a 14 MeV proton beam and degrader foils. In their measurements, proton energy uncertainty was high. In this study a reduction in energy uncertainty on cross-section measurement was done along with a new measurement around 2.6 MeV energy. The Molecular deposition technique has been used to prepare $^{144}$Sm$_2$O$_3$ (67\% enriched) targets on pure Al (99.45\%) backing~\cite{bar2022preparation}. The experiment was performed at the K130 cyclotron at VECC, Kolkata. As the lowest available beam energy is 7 MeV, stack foil technique was used to beam energy. The irradiated targets were counted using HPGe detectors. The (p, $\gamma$) reaction cross-sections were analysed using the Hauser-Feshbach statistical model and satisfactory results were obtained.
\section{Experiment}

The experiment was performed at VECC, Kolkata using the stacked foil activation technique. It was then followed by an offline $\gamma-$ray spectroscopy measurement. Several stacks were individually bombarded with 7 MeV proton beam. Cross-section measurements were carried out for proton energies between 2.6 and 6.8 MeV. The proton beam energy was decreased using degrader foils. $^{144}$Sm targets were irradiated for a few hours to couple of days with a beam current of $\sim$1 $\mu$A. Copper foils placed in some stacks, were used to determine the beam intensity. Details of the experimental procedure and data analysis are described below.
\subsection{Target preparation}

Irradiation targets have been made from 67\% isotopically enriched $^{144}$Sm$_2$O$_3$ powder using molecular deposition technique~\cite{parker1964molecular,bar2022preparation}. 10 mg oxide power was dissolved in $\sim$250 $\mu$l HNO$_3$. This solution was then dried under an infrared (IR) lamp for approximately 15 minutes. After the solution completely dried out, $\sim$500 $\mu$l of deionized (DI) water was poured in and mixed well so that all Sm(NO$_3$) molecules get dissolved. This solution was again placed under the IR lamp to dry, this process was repeated for a couple of times. Finally the target solution was prepared by adding 2 ml DI water with the evaporated target sample residue. Some 120 $\mu$l of the solution was added to a cylindrical deposition chamber having $\sim$25 ml isopropyl alcohol in it (Fig. 1). A circular end platinum wire (1 mm diameter) and 25 $\mu$m pure (99.45\%) Al foil connected to positive and ground terminals of the power supply. Deposition was carried out for approximately 45$-$60 minutes with anode voltage of +500 V. Targets of thickness between 100$-$350 $\mu$g/cm$^2$ was prepared.
\subsection{Preparation of the target stacks and experimental setup}
\begin{figure}
  \centering
  \begin{subfigure}[b]{0.48\linewidth}
    \includegraphics[clip, trim=5.0cm 0.0cm 4.0cm 0.0cm,width=\linewidth]{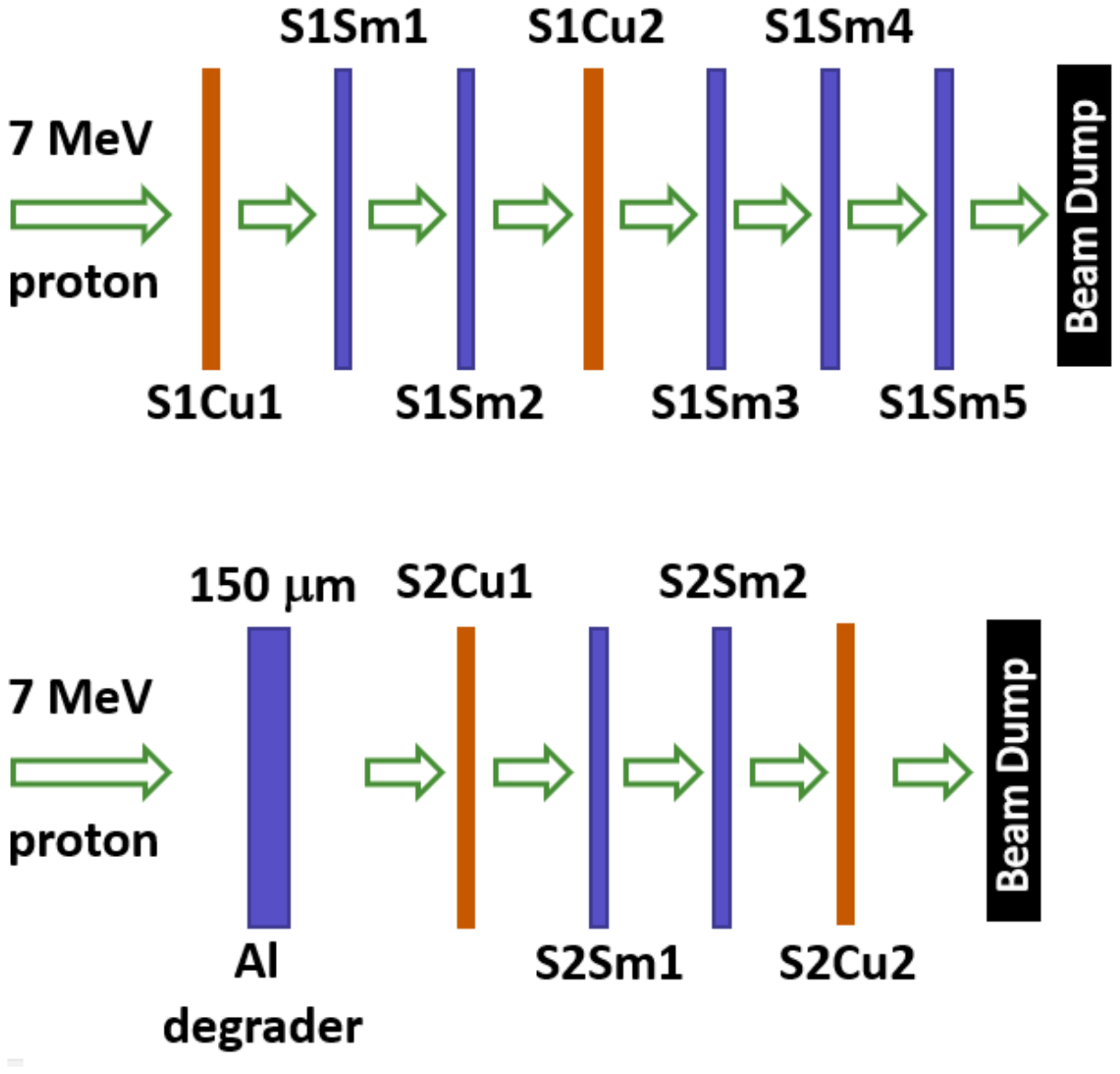}
    \caption{Target and monitor foil positions for Stack 1 (S1) and Stack 2 (S2).}  
  \end{subfigure}
  \begin{subfigure}[b]{0.35\linewidth}
    \includegraphics[clip, trim=0.0cm 0.0cm 0.0cm 0.0cm,width=\linewidth]{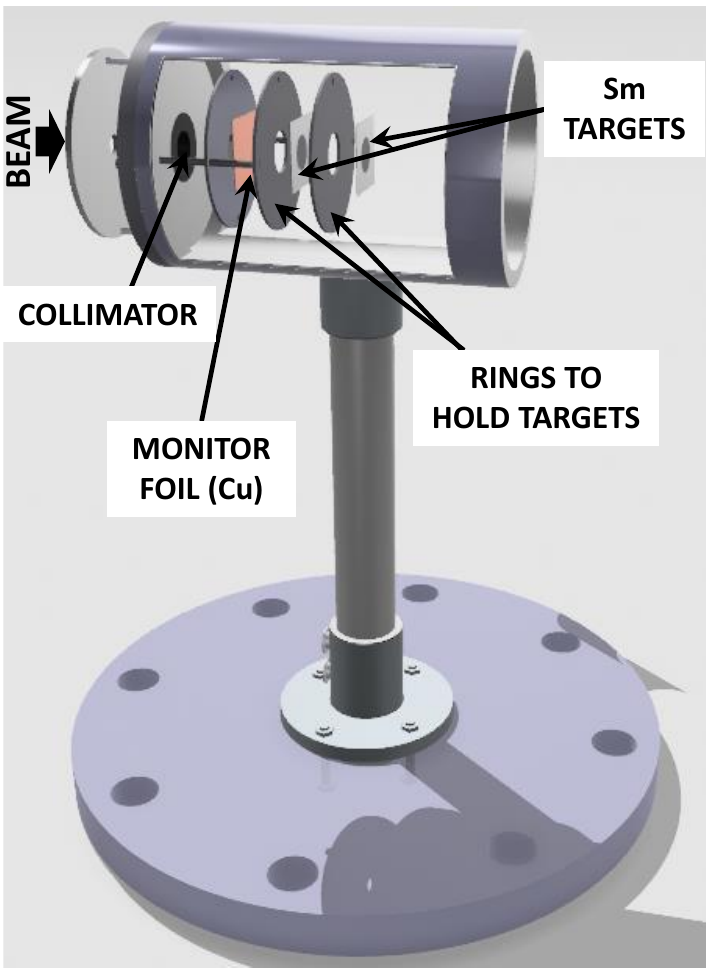}
    \caption{Schematic diagram of the setup.}
  \end{subfigure}
  \caption{Target stack setup and target positions for multiple target irradiation between energy 4.2$-$6.8 MeV.}
  \label{stack_setup}
\end{figure}

\begin{figure}
  \centering
    \includegraphics[clip, trim=0.0cm 0.0cm 0.0cm 0.0cm,width=0.8\linewidth]{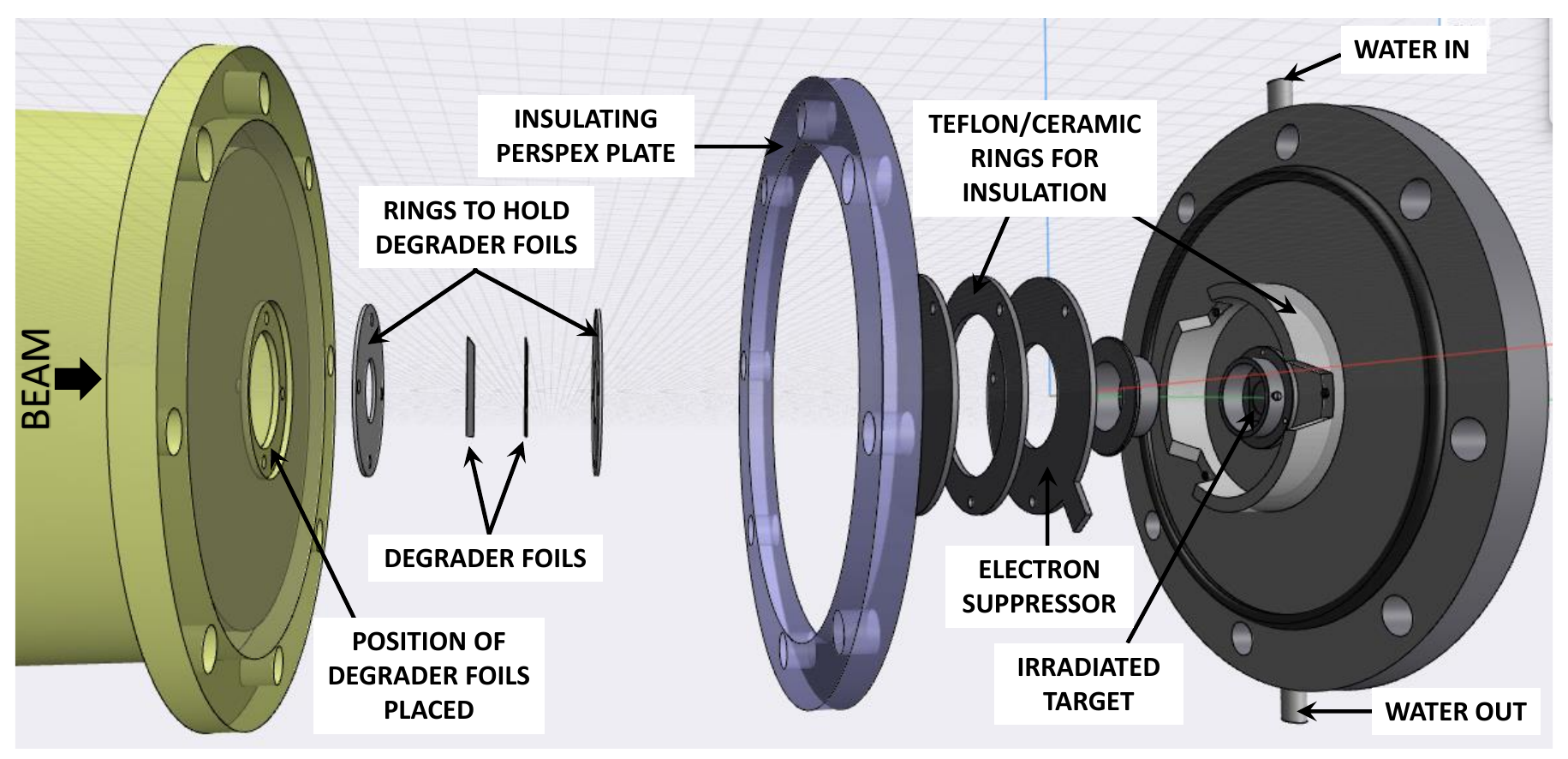}
    \caption{Single target irradiation setup with electron suppressor. This setup is used to irradiate $^{144}$Sm targets for energy below 4.2 MeV.} 
    \label{setup3} 
  \end{figure}

Two stacks of targets (Figure~\ref{stack_setup}) have been used to measure cross-sections between energy 4.2$-$6.8 MeV. First stack (S1) targets were irradiated for energies between 5.1 and 6.8 MeV (five energies). The second stack (S2) consisted of several pure Al foils of accumulated thickness of 150 $\mu$m placed at the beginning to allow irradiation of $^{144}$Sm targets at energies 4.7 and 4.2 MeV. Two copper foils were also placed at different positions to monitor the beam intensity. At the end of each stack, a beam dump was placed. Schematic picture of actual setup is shown in Figure~\ref{stack_setup}. Target irradiation below 4.2 MeV was achieved using single $^{144}$Sm target with Al degrader foils of varying thickness before it (Figure~\ref{setup3}). Target holding flange was cooled using low conductivity water during the irradiation. A separate plate was placed as an electron suppressor in front with $-$500 V. Beam current was recorded every minute to include any current variation during the single target irradiation. This setup has been used to measure cross-section for energies between 2.6 and 4.1 MeV proton energy (five energies). 

\subsection{Irradiation energy and beam current determination}

\begin{figure}
  \centering
    \includegraphics[clip, trim=0.0cm 0.0cm 0.0cm 0.0cm,width=0.8\linewidth]{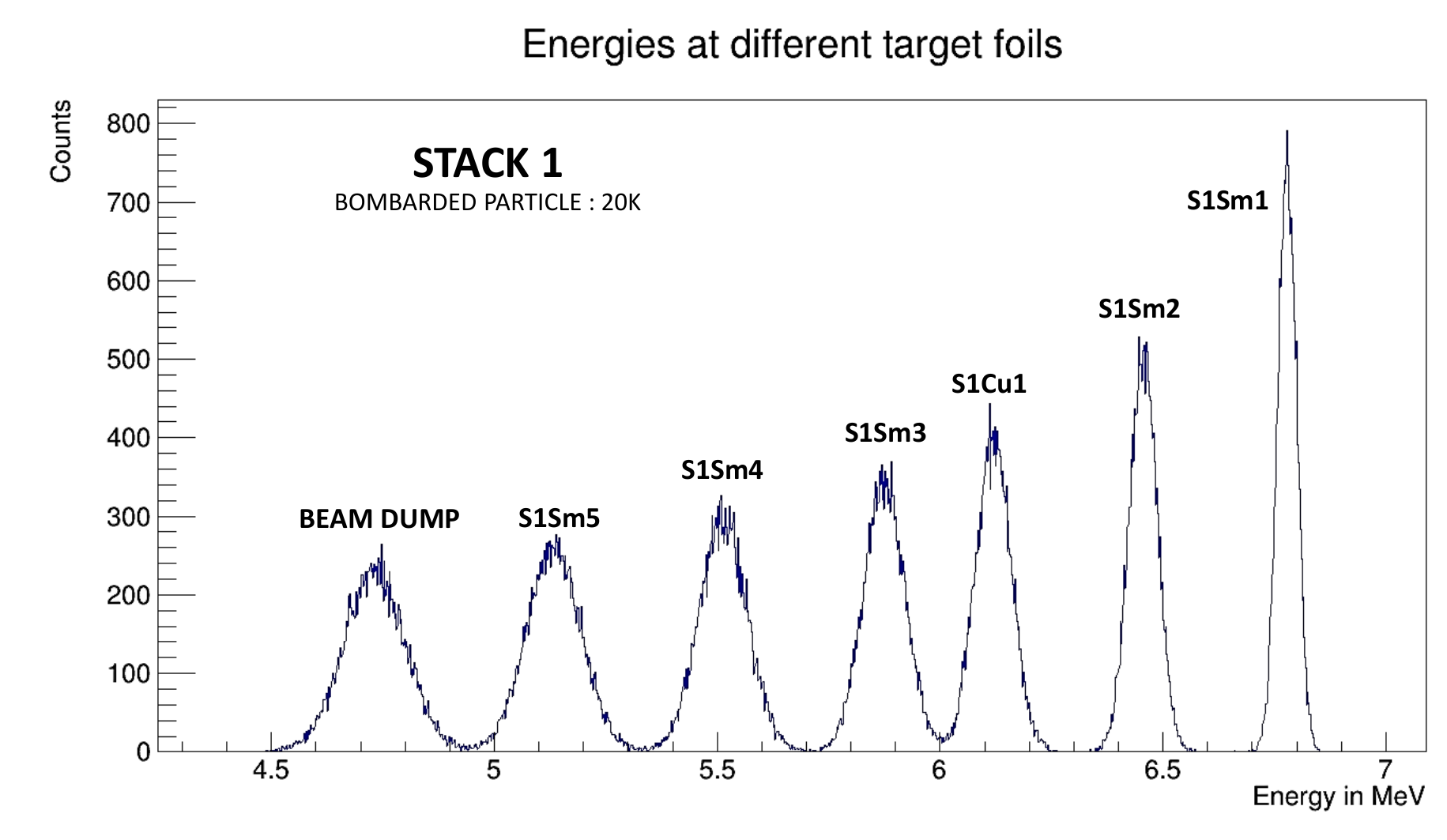}
    \caption{Geant4 simulation for estimating energy at each foil of Stack 1 (S1).} 
    \label{geant4} 
  \end{figure}

Aluminium backing present in the $^{144}$Sm targets acts as an energy degrader for successive $^{144}$Sm targets. Degraded proton energy with energy straggling (1$\sigma$) were estimated using LISE++ code~\cite{tarasov2016lise++} and GEANT4 simulation package~\cite{apostolakis2007overview}. A result obtained from GEANT4 simulation is shown in Figure~\ref{geant4}. Results from GEANT4 and LISE++ are in agreement with each other within 5\%.

Simulation result shows (Figure~\ref{geant4}) that as beam passes through multiple foils, more and more energy degradation happen. This also brings some uncertainty to the beam energy. Energy equal to the 1$\sigma$ value was considered as the energy straggling in the irradiating beam. 
S1 and S2 stacks (marked in Figure~\ref{stack_setup})were are irradiated for approximately 17 and 39 hours respectively. Beam intensity for these bombardments were calculated from monitor copper foil (7 $\mu$m) placed between target foils at different positions using equation~\ref{cross_eq2}. $^{65}$Cu(p, n)$^{65}$Zn reaction cross-section data, which was used for beam current estimation, has been taken from ~\cite{generalov2017activation}. Single targets (4.1 MeV and below) were irradiated between 17$-$45 hours. Bombardment time was increased as the cross-section decreases drastically with lower irradiation energy. Beam current for these targets were measured from the end flange used as the target holder. 
\subsection{Measurement of $\gamma-$ray activity}

\begin{figure}
  \centering
    \includegraphics[clip, trim=2.0cm 5.0cm 2.0cm 5.0cm,width=0.8\linewidth]{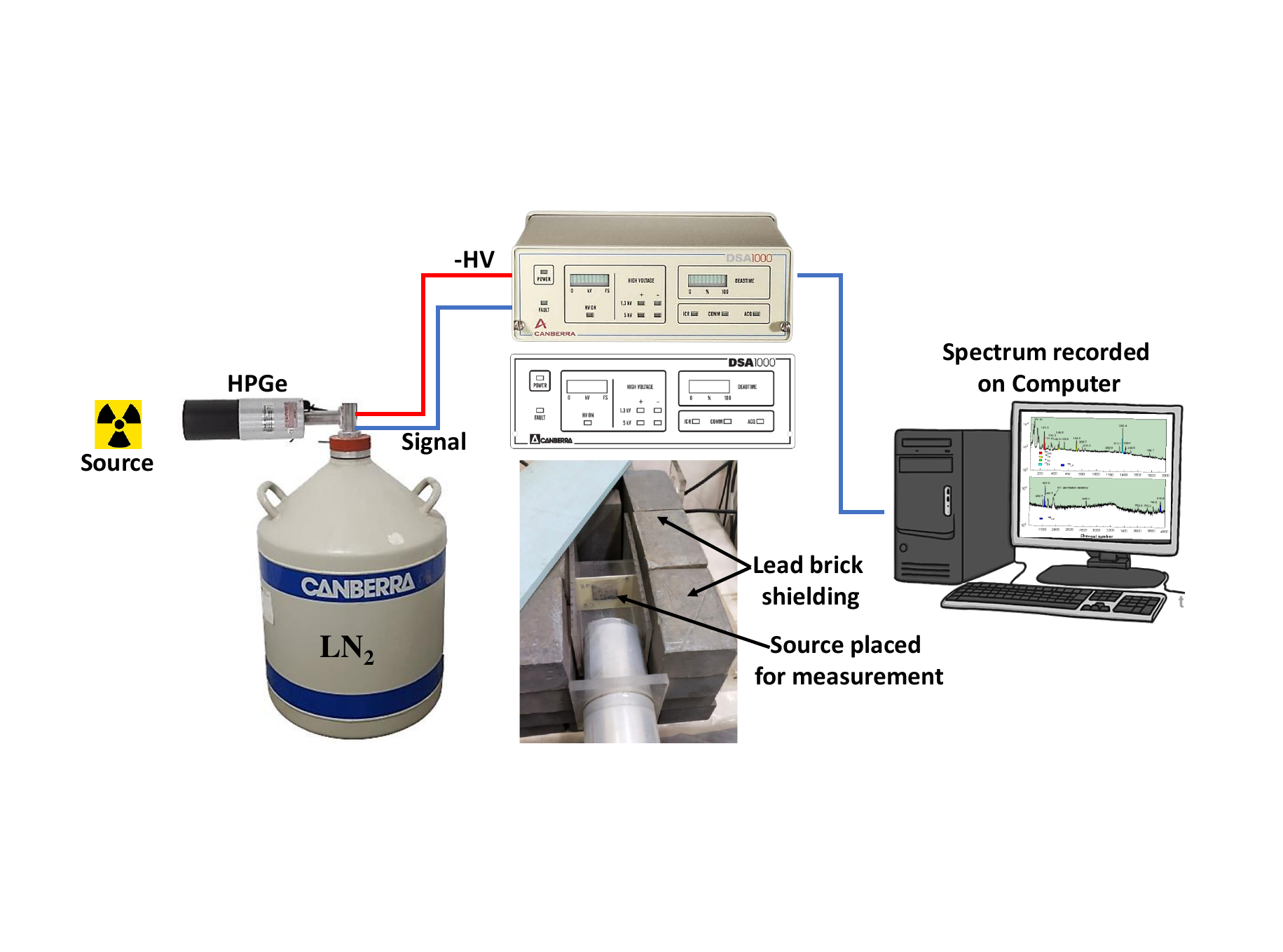}
    \caption{Schematic diagram of detection setup used for $\gamma-$ spectroscopy measurement.} 
    \label{detsetup} 
  \end{figure}

After the irradiation, stacks (S1 and S2) were cooled for 24$-$48 hours (due to higher activity from multiple Sm and Cu foils) and targets irradiated individually were rested for 1$-$2 hours before measuring $\gamma-$activity. The $\gamma-$activity was measured using a CANBERRA high-purity germanium (HPGe) detector having 40\% relative efficiency and $\sim$1.8 keV energy resolution at 1.33 MeV $\gamma-$energy. Detector was shielded by 7.5 cm thick lead bricks and the data acquisition was carried out using a full featured 16K channel integrated multichannel analyzer (MCA) based on digital signal processing technology, CANBERRA DSA1000 and a spectroscopy software suite called GENIE. Figure~\ref{detsetup} shows a schematic setup diagram of the complete detection system. 

\subsection{HPGe detector efficiency calibration}

A standard $^{152}$Eu (T$_{1/2}$=13.517$\pm$0.009 years) point source of known activity (A${_o}$=3.908 $\times$10$^4$ Bq as on 17 May 1982) was used to measure the absolute photo-peak efficiency of the HPGe detector at different $\gamma-$energies. The detector efficiency for the point source was calculated using~\cite{PhysRevC.95.024619},
\begin{equation}
\epsilon_p=\frac{C_\gamma K_C}{I_\gamma A_o e^{-\lambda t} t_{meas}}
\label{eff_eq}
\end{equation}
where C$_\gamma$ is the count under the peak of energy $\gamma$, A$_o$ is the $^{152}$Eu source activity at the time of production, t is the time elapsed between Eu-source production and start of counting time. I$_\gamma$ and $\lambda$ ($=\frac{0.693}{T_{1/2}}$, $T_{1/2}$ being half-life of nuclei) represents decay intensity and decay constant respectively, t$_{meas}$ is the counting time and K$_C$ is the correction factor to counter any summing effect.

Irradiated targets were placed very close (12.5 mm) to the detector surface due to low activity of $^{144}$Sm(p, $\gamma$)$^{145}$Eu reaction at measured energies. Therefore, standard source for calibration had to be placed at the same distance. Such a close geometry gives rise to the possibility of co-incidence summing. To obtain appropriate detector efficiency, correction due to summing effect need to be incorporated, which is discussed in the next section.

\subsection{Co-incidence summing effect}

The co-incidence summing occurs in a $\gamma-$spectroscopy when two $\gamma-$rays emitted in a cascade, enter the detector active volume within the resolving time or data processing time of the detector. The detector in such situations cannot distinguish between the two and count them as a single event, with the energy equal to the sum of the two individual photon energies. This is called $``$\textit{true co-incidence}" or $``$\textit{cascade co-incidence}" summing. This event can cause decrease (summing out) or increase (summing in) in counts for $\gamma-$ray of interest.\\
For any detector and source configuration there is always some degree of summing effects present. This effect is more prominent when the source is very close to the detector surface. This effect can be incorporated in the efficiency calculation by introducing a correction factor (K$_C$) in the efficiency calculation (Eq.~\ref{eff_eq}). This correction factor has been calculated using a Monte Carlo code called EFFTRAN~\cite{vidmar2011calculation,rameback2021validation}. The detailed specification of the detector and source dimensions were provided as an input to the program.\\
Another efficiency correction required due to the finite geometry of the sample targets unlike the standard source which is point like used for efficiency calibration. Irradiated targets have a $\sim$6 mm diameter active area due to bombardment. The efficiency values obtained from the point source standard after the summing corrections are then used to find the efficiency for the finite dimension samples using EFFTRAN code. Obtained results are shown in Table~\ref{tab_eff}.
\begin{table}
\centering
\caption{\label{tab_eff}Efficiency,Correction for summing effect($K_C$), point source efficiency($\varepsilon_p$), Corrected efficiency for Extended Sample ($\varepsilon_s$)}
\footnotesize
\begin{tabular}{@{}ccccccc}
\br
$\gamma-$energy&Intensity(\%)&Area under peak(C$_\gamma$)&$K_C$&$\varepsilon_p$& $\varepsilon_s$\\
\mr
121.8&28.53$\pm$0.16&467836(0.19\%)&1.131&0.1293&0.1254(0.93\%)\\
\mr
244.7&7.55$\pm$0.04&79367(0.48\%)&1.193&0.0874&0.0847(1.00\%)\\
\mr
344.3&26.59$\pm$0.02&262917(0.25\%)&1.088&0.0750&0.0727(0.77\%)\\
\mr
411.1&2.237$\pm$0.013&15032(1.07\%)&1.223&0.0573&0.0556(1.39\%)\\
\mr
444.0&3.125$\pm$0.018&21120(0.84\%)&1.170&0.0551&0.0534(1.22\%)\\
\mr
778.9&12.93$\pm$0.08&63148(0.51\%)&1.129&0.0384&0.0372(1.07\%)\\
\mr
964.1&14.51$\pm$0.07&60747(0.52\%)&1.092&0.0319&0.0309(1.00\%)\\
\mr
1112.1&13.67$\pm$0.08&53163(0.55\%)&1.049&0.0284&0.0276(1.07\%)\\
\mr
1408.0&20.87$\pm$0.09&65180(0.51\%)&1.068&0.0233&0.0226(1.00\%)\\
\br
\end{tabular}
\end{table}
\begin{figure}[h]
\centering
\includegraphics[clip, trim=0.0cm 0.0cm 0.0cm 0.0cm,width=0.6\textwidth]{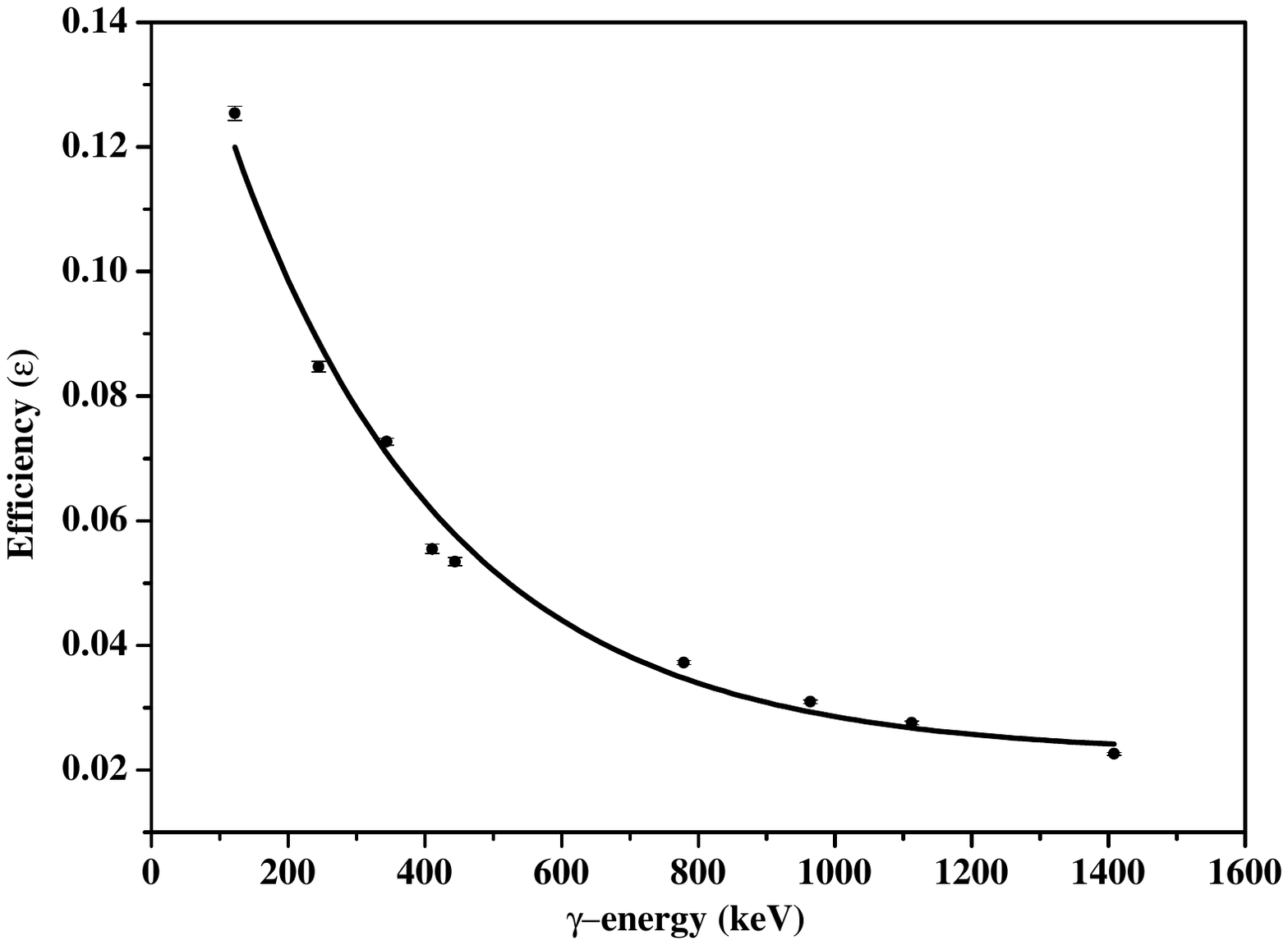}
\caption{Efficiency curve.}
\label{eff_curve}
\end{figure}

\subsection{Detector efficiency curve}

After incorporating corrections due to summing effect and finite dimension of samples, Table~\ref{tab_eff} used to find a relation between detector efficiency and specific gamma energy. Figure~\ref{eff_curve} shows the relation and fitting was done using the Eq.~\ref{eff_fit}.
\begin{equation}
\varepsilon_s (E_\gamma)=\varepsilon_1 e^{-E_\gamma/E_0}+\varepsilon_0
\label{eff_fit}
\end{equation}
The fitting parameters ($\varepsilon_1$, E$_o$ and $\varepsilon_0$) obtained from detector efficiencies for standard source are listed in Table~\ref{eff_fitTab}.
\begin{table}
\centering
\caption{\label{eff_fitTab}Efficiency curve fitting parameters}
\footnotesize
\begin{tabular}{@{}ccc}
\br
Parameter&Fitted value&Uncertainty\\
\mr
$\varepsilon_1$&0.143&0.012\\
\mr
$\varepsilon_0$&0.023&0.002\\
\mr
$E_0$&317.384&34.985\\
\br
\end{tabular}
\end{table}
This relation was used to determine detector efficiency at specific gamma energy coming from the activated sample targets.

\section{Data Analysis}

\begin{figure}
  \centering
    \includegraphics[clip, trim=0.0cm 0.0cm 0.0cm 0.0cm,width=0.9\linewidth]{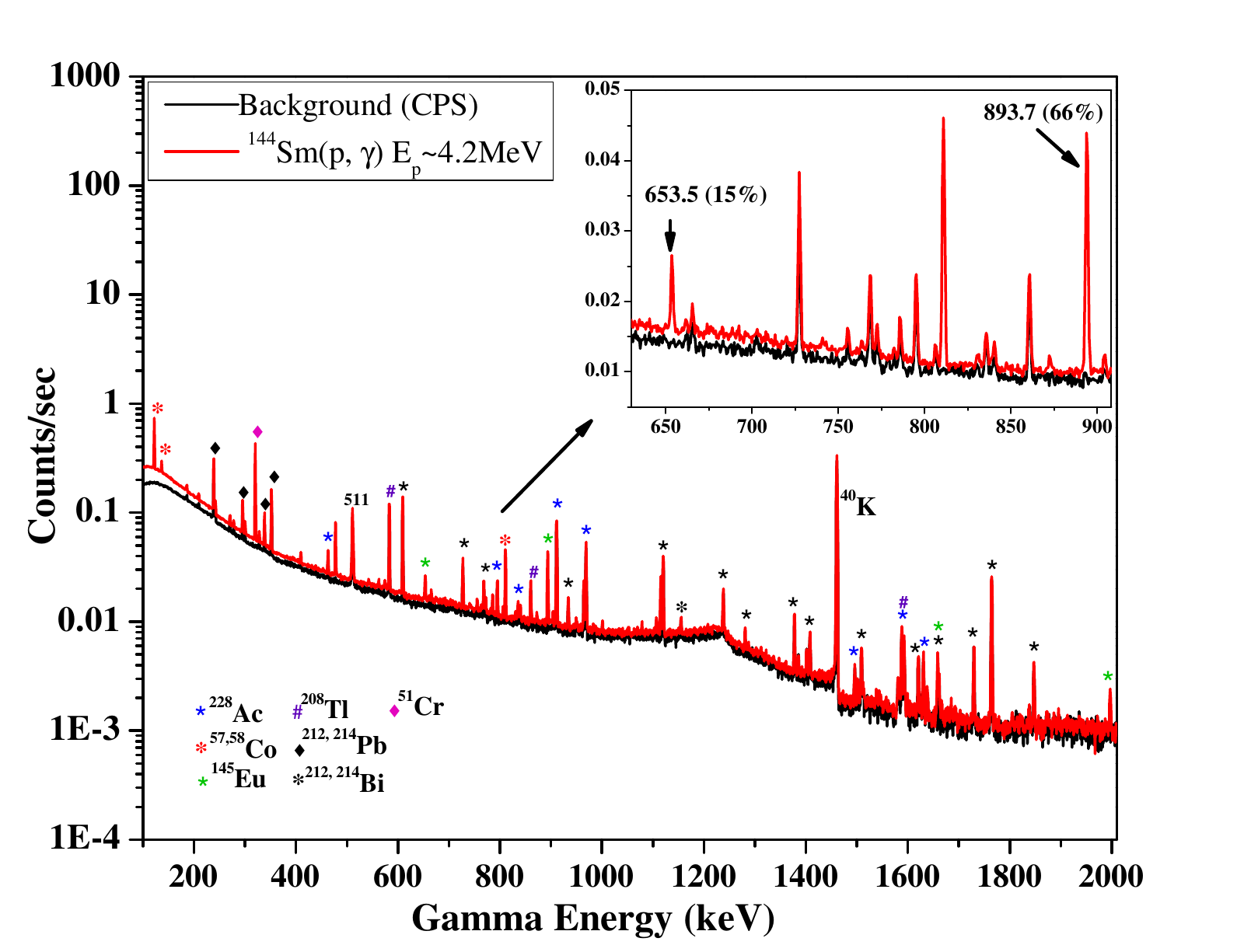}
    \caption{Typical gamma spectroscopy from a $^{144}$Sm target is shown in red and black represents the background in the counting area. Prominent $\gamma-$rays present in the spectrum are marked.} 
    \label{spec} 
  \end{figure}

Cross-section of $^{144}$Sm(p,$\gamma$)$^{145}$Eu (Q-value$=$3.315 MeV) reaction was measured for eleven different proton energies. Monitor foil (natural Cu) was bombarded at four different energies to find beam intensity. All irradiated energies and corresponding energy straggling (1$\sigma$) due to energy degradation by foils placed upfront have been listed in Table~\ref{irr_energy}.
\begin{table}
\centering
\caption{\label{irr_energy}Irradiated energies of $^{144}$Sm and natural copper and respective energy uncertainty (1$\sigma$) as calculated with LISE++. Accelerator beam energy uncertainty of 0.5\% (FWHM) is also included.}
\footnotesize
\begin{tabular}{@{}cccc}
\br
&Irradiated energy&Irradiated Target&Energy uncertainty (1$\sigma$)\\&(MeV)&& (MeV)\\
\mr
Stack~1&7.00&Nat. Cu&$\pm$ 0.02\\
&6.78&$^{144}$Sm&$\pm$ 0.02 \\
&6.46&$^{144}$Sm&$\pm$ 0.03 \\
&6.13&Nat. Cu&$\pm$ 0.04 \\
&5.88&$^{144}$Sm&$\pm$ 0.05 \\
&5.52&$^{144}$Sm&$\pm$ 0.05 \\
&5.13&$^{144}$Sm&$\pm$ 0.06 \\
\mr
Stack~2&4.97&Nat. Cu&$\pm$ 0.07\\
&4.69&$^{144}$Sm&$\pm$ 0.07 \\
&4.26&$^{144}$Sm&$\pm$ 0.08 \\
&3.79&Nat. Cu&$\pm$ 0.08 \\
\mr
Single target&4.11&$^{144}$Sm&$\pm$ 0.09\\
&3.68&$^{144}$Sm&$\pm$ 0.10 \\
&3.17&$^{144}$Sm&$\pm$ 0.11 \\
&2.59&$^{144}$Sm&$\pm$ 0.13 \\
\br
\end{tabular}
\end{table}
$\gamma-$detection measurement of irradiated targets was carried out for 15 minutes to couple of days to accumulate sufficient statistics. A typical $\gamma-$ray spectrum for a $^{144}$Sm target bombarded at $\sim$4.2 MeV along with background is shown in Figure~\ref{spec}.
The bombarded target activities were calculated using the Eq.~\ref{cross_eq1},
\begin{equation}
A= \frac{\lambda C_\gamma}{\varepsilon_d I_\gamma e^{-\lambda t_{cool}} (1-e^{-\lambda t_{meas}})}
\label{cross_eq1}
\end{equation}
where A is the activity at the end of irradiation, $\varepsilon_d$ stands for detector efficiency, C$_\gamma$ is the count under the peak, t$_{cool}$ is the time between end of irradiation and start of counting, and t$_{meas}$ is the counting time. $\gamma-$rays listed in Table~\ref{gammalist} were measured for analyses. Bombarded proton flux for S1 and S2 stacks was estimated from the $^{65}$Zn activity using,
\begin{equation}
\phi_b= \frac{A}{\sigma N_t (1-e^{-\lambda t_{irr}})}
\label{cross_eq2}
\end{equation}
where $\phi_b$ denotes proton flux (1/s), $\sigma$ is the cross-section (cm$^2$), N$_t$ is the target atoms/cm$^2$ and t$_{irr}$ is the irradiation time. The proton flux was estimated for S1 and S2 stacks using Eq.~\ref{cross_eq2}.
\begin{table}
\centering
\caption{\label{gammalist}List of detected $\gamma-$rays from different nuclei with half-life and $\gamma-$ray intensity}
\footnotesize
\begin{tabular}{@{}ccc}
\br
Produced nuclei&Half$-$life (T$_{1/2}$)~\cite{nndc}&Detected $\gamma-$rays \\&& with intensity~\cite{nndc}\\
\mr
$^{145}$Eu&5.93(4) days&893 (66\%)\\&&653.5(15\%)\\
\mr
$^{65}$Zn&243.93(9) days&1115.5(50\%) \\

\br
\end{tabular}\\
\end{table} 
Combining equation~\ref{cross_eq1} and~\ref{cross_eq2}, activation formula for cross-section measurement can be obtained as,
\begin{equation}
\sigma = \frac{C_\gamma~\lambda}{\phi_b~N_t~I_\gamma~\varepsilon_d~(1-e^{-\lambda t_{irr}}) (e^{-\lambda t_{cool}} - e^{-\lambda(t_{cool} + t_{meas})})}
\label{cross_eq3}
\end{equation}
Peak count of $\gamma-$energies were calculated by subtracting neighbouring background counts of lower and higher energy sides. These analyses were done in CERN ROOT Data Analysis Framework~\cite{brun1997root}. The uncertainty in the measured cross-sections increases from 25\% to 45\% with the reduction of proton energy. This was calculated by considering uncertainties from all possible sources like $\gamma-$intensity, decay constant, target thickness, standard source activity and also from the statistical and fitting of peak area and detector efficiency curve. Uncertainty in the $^{65}$Cu(p, n) reaction cross-section due to the energy loss inside the copper foil is also accounted in the proton current estimation for Stack 1 and Stack 2 irradiation energies.

\section{Theoretical models}
\begin{figure}[h]
\centering
\includegraphics[clip, trim=0.0cm 0.0cm 0.0cm 0.0cm,width=0.75\textwidth]{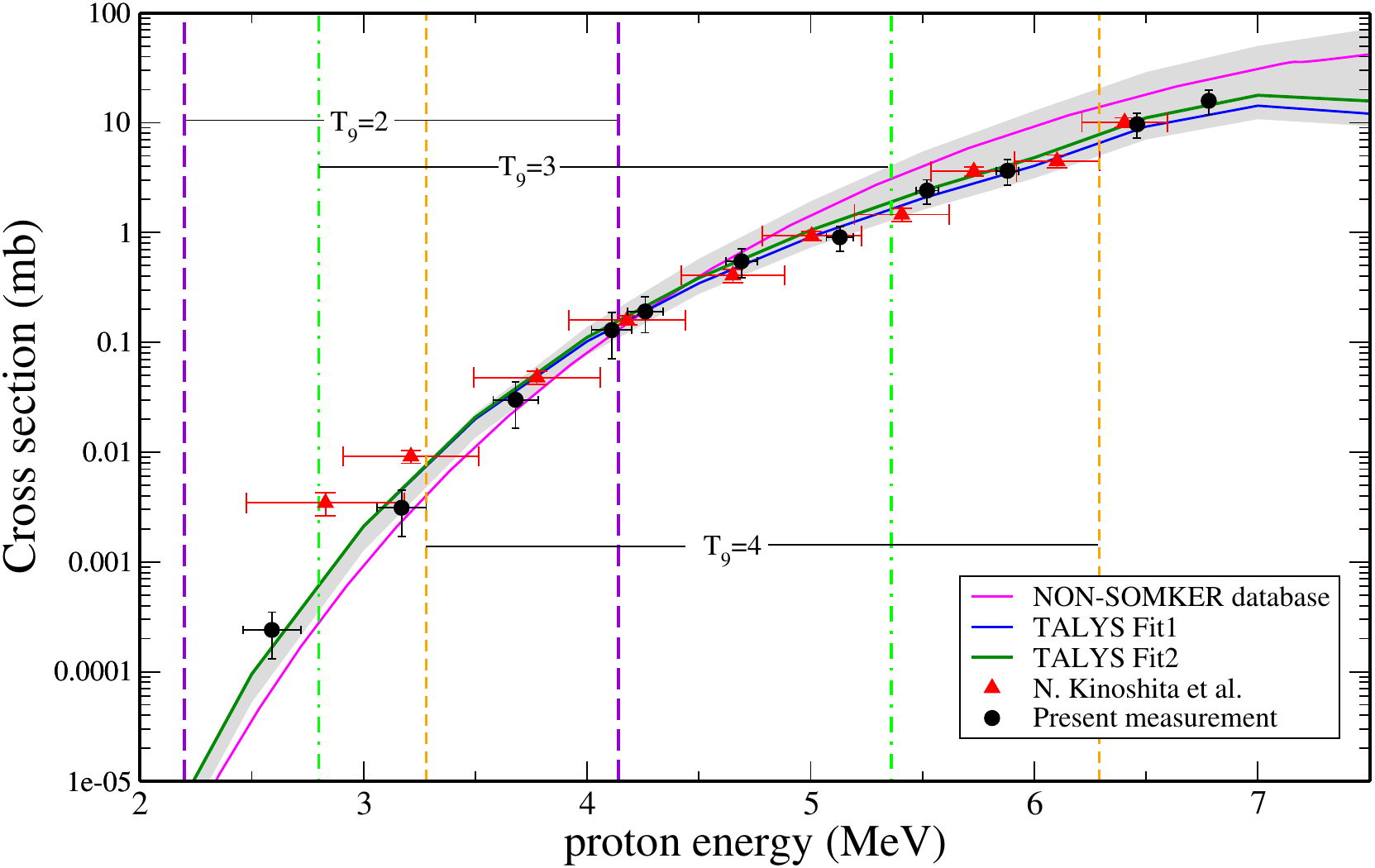}
\caption{The experimental cross-sections and Hauser-Feshbach calculations obtained from TALYS 1.96 and NON-SMOKER database are presented. Grey shaded area denotes the theoretically calculated cross-section from TALYS by varying all OPM, NLD and $\gamma-$SF combinations. TALYS Fit1 obtained with OPM 1, ldmodel 5 and $\gamma-$SF 1 whereas TALYS Fit2 uses OPM 1, ldmodel 5 and $\gamma-$SF 3. Gamow window corresponding to different stellar temperatures are also marked. }
\label{cross_fig}
\end{figure}

The measured $^{144}$Sm(p, $\gamma$)$^{145}$Eu reaction cross-sections (Figure~\ref{cross_fig}) are compared with the values obtained from statistical Hauser-Feshbach model code TALYS 1.96~\cite{koning2023talys}. Gamow windows corresponding to each stellar temperature are also marked. Theoretical prediction of cross-section in TALYS was done by implementing a total of 216 combinations of optical potential model (OPM), nuclear level density (NLD) and $\gamma-$ray strength function ($\gamma-$SF). All the models used during calculation are listed in Table~\ref{Model_list}.

\begin{table}
\centering
\caption{\label{Model_list}Different models used during TALYS calculation. A total of 216 combinations from these models have been tried to find the cross-section predictions.}
\footnotesize
\begin{tabular}{@{}lll}
\br
Parameter&Models&Details\\
\mr
Optical potential& OPM 1& Koning and Delaroche local potential~\cite{koning2003local}\\
& OPM 2& Koning and Delaroche global potential~\cite{koning2003local}\\
& OPM 3& Koning-Delaroche local dispersive potential\\
& OPM 4& \parbox[t]{10cm}{Jeukenne-Lejeune-Mahaux (JLM) OPM calculations, a semi-microscopic optical model~\cite{bauge2001lane}}\\
\mr
Nuclear level density&ldmodel 1& Constant temperature and Fermi gas model (CTM)~\cite{gilbert1965composite,ericson1960statistical}\\
&ldmodel 2& back-shifted Fermi-gas model (BFM)~\cite{dilg1973level}\\
&ldmodel 3& Generalised Superfluid model(GSM)~\cite{ignatyuk1979kk,ignatyuk1993density}\\
&ldmodel 4& \parbox[t]{10cm}{Skyrme-Hartree-Fock-Bogoluybov level densities from numerical tables (microscopic model)~\cite{goriely2006microscopic}}\\
&ldmodel 5& \parbox[t]{10cm}{Gogny-Hartree-Fock-Bogoluybov level densities from numerical tables (microscopic model)~\cite{goriely2006microscopic}}\\
&ldmodel 6& \parbox[t]{10cm}{Temperature-dependent Gogny-Hartree-Fock-Bogoluybov level densities from numerical tables (microscopic model)~\cite{hilaire2012temperature}}\\
\mr
$\gamma-$ray strength function&$\gamma-$SF 1& Kopecky-Uhl generalized Lorentzian~\cite{kopecky1990test}\\
&$\gamma-$SF 2& Brink-Axel Lorentzian~\cite{brink1957individual,axel1962electric}\\
&$\gamma-$SF 3& Hartree-Fock BCS tables~\cite{goriely2002large}\\
&$\gamma-$SF 4& Hartree-Fock-Bogoliubov (HFB) tables~\cite{goriely2004microscopic}\\
&$\gamma-$SF 5& Goriely’s hybrid model~\cite{goriely1998radiative}\\
&$\gamma-$SF 6& Goriely T-dependent HFB~\cite{goriely2004microscopic}\\
&$\gamma-$SF 7& Temperature-dependent Relativistic Mean Field (RMF) model~\cite{daoutidis2012large}\\
&$\gamma-$SF 8& \parbox[t]{10cm}{Gogny-Hartree-Fock-Bogoliubov model with the quasiparticle random phase approximation (QRPA)~\cite{goriely2018gogny}}\\
&$\gamma-$SF 9& \parbox[t]{10cm}{Simplified Modified Lorentzian (SMLO) by Stephane Goriely and Vladimir Plujko~\cite{plujko2019description}}\\
\br
\end{tabular}
\end{table}
The grey shaded area marked in Figure~\ref{cross_fig} indicate the region of calculated values obtained from TALYS by variation of all input parameter combinations. Boundary of this gray region represents maximum and minimum cross-section value theoretically obtained for each energy. It was found that combination of OPM 1(default), 2, 3; ldmodel 5, 6 with $\gamma-$SF 3 gives similar and best fitting of the measured data. Although two different set of parameters have been chosen for comparison purpose. In Figure~\ref{cross_fig}, TALYS Fit1 values are obtained for OPM 1, ldmodel 5 and $\gamma-$SF 1 whereas TALYS Fit2 is for OPM 1, ldmodel 5 and $\gamma-$SF 3. Further theoretical calculation of reaction rate was done separately for parameters from TALYS Fit1 and TALYS Fit2.

\subsection{Sensitivity}

The sensitivity study is crucial to understand the effect of input parameters in nuclear models to the cross-section value. The sensitivity can be defined as~\cite{rauscher2012sensitivity},

\begin{equation}
S_q^\sigma= \frac{d\sigma/\sigma_{old}}{dq/q_{old}}
\label{sens}
\end{equation}
where d$\sigma=\sigma_{new}-\sigma_{old}$ and dq$=$q$_{new} -$ q$_{old}$. $S_q^\sigma=0$ means no change in $\sigma$ value even if there is any change in q-value. Whereas $S_q^\sigma=1$ implies change in $\sigma$ by the same manner as in q.
\begin{figure}[h]
\centering
\includegraphics[clip, trim=0.0cm 0.0cm 0.0cm 0.0cm,width=0.7\textwidth]{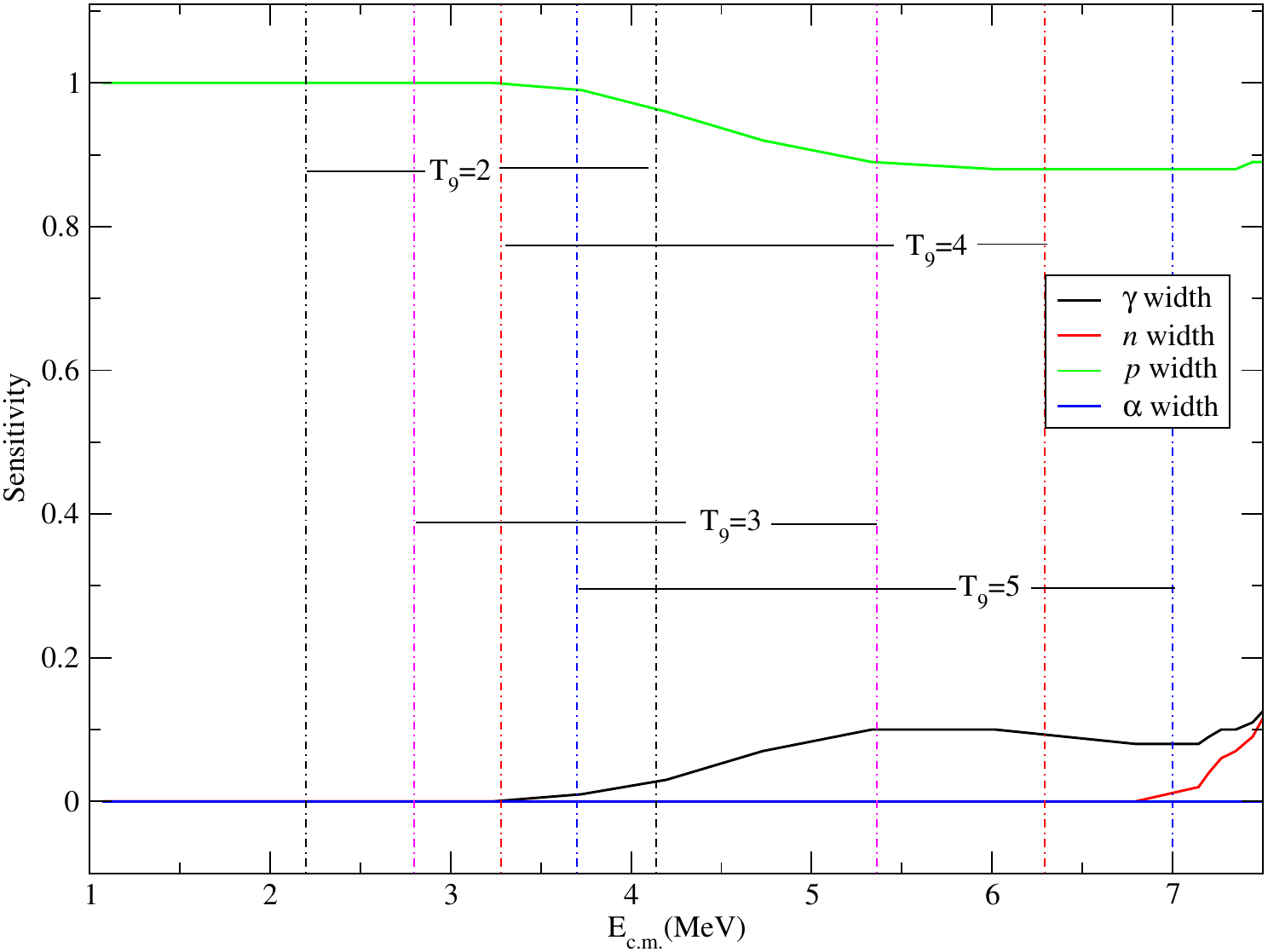}
\caption{Sensitivity plot.}
\label{sens_fig}
\end{figure}

Figure~\ref{sens_fig} shows that the sensitivity of $^{144}$Sm(p, $\gamma$) reaction cross-section against the proton, neutron, neutron, alpha and $\gamma-$width. It is clearly evident from the plot that proton width is the dominant factor while determining the $^{144}$Sm(p, $\gamma$) cross-section in the astrophysically important energy regime. At energies $>$4 MeV, $\gamma-$width and energies higher than the $^{144}$Sm(p, n) threshold ($=7.179$ MeV) neutron width contributes to the cross-section as well. In the present measurement, the theoretically obtained cross-sections agree well without any correction of the proton width at 2.6-6.8 MeV. Gamow window related to stellar temperature are also mentioned.

\subsection{S-factor calculation}

\begin{figure}[h]
\centering
\includegraphics[clip, trim=0.0cm 0.0cm 0.0cm 0.0cm,width=0.7\textwidth]{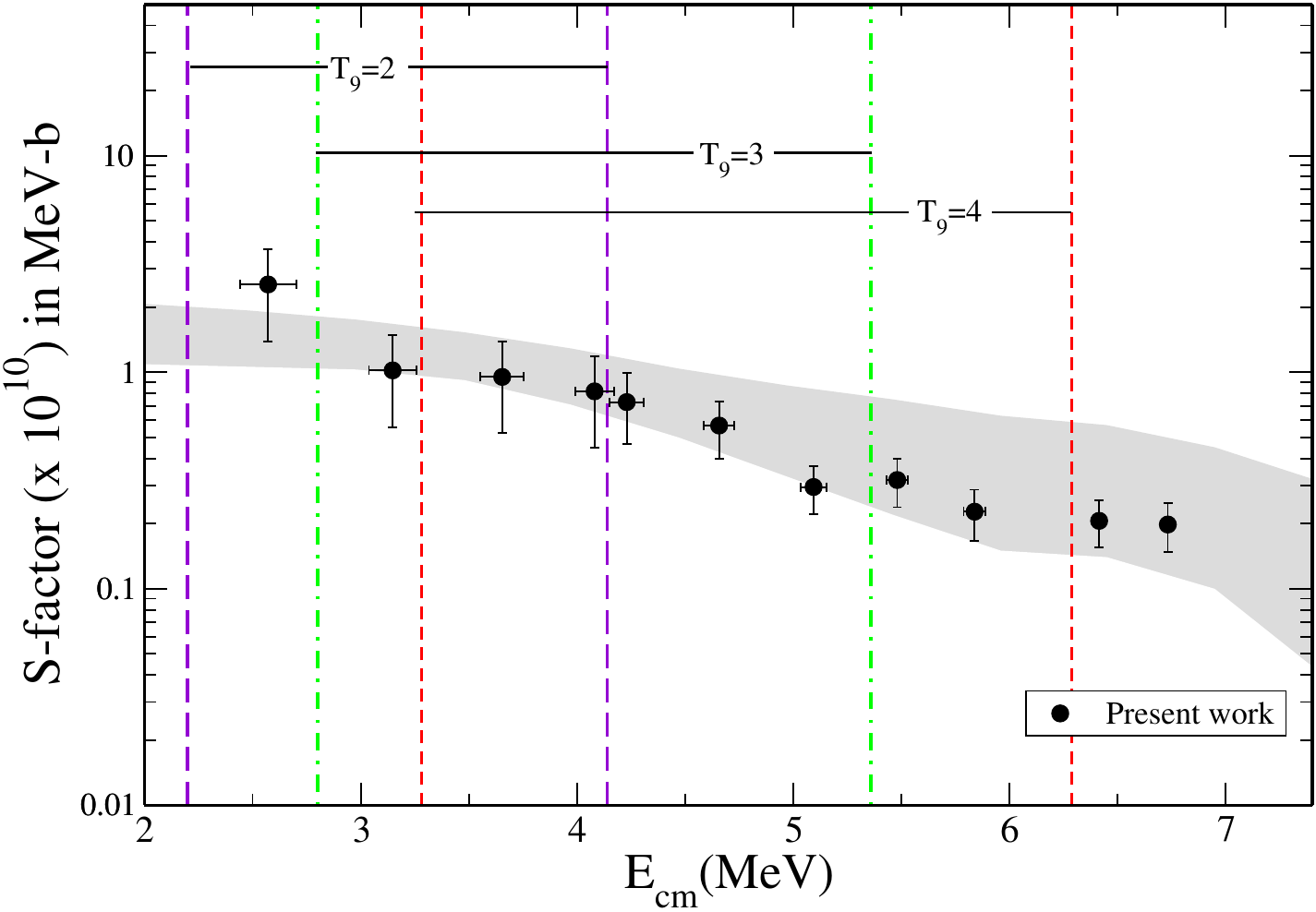}
\caption{The astrophysical $S-$factors calculated from experimental cross-section of $^{144}$Sm(p, $\gamma$)$^{145}$Eu reaction and theoretical predictions are shown with the grey shaded region.}
\label{sfactor_fig}
\end{figure}

The reaction cross-section of $^{144}$Sm(p, $\gamma$) has been measured in the center of mass energies between 2.6 and 6.8 MeV. Astrophysical S-factor has been calculated for the above reaction in the measured energy range. S-factor is defined as~\cite{iliadis2015nuclear},
\begin{equation}
S(E_{cm})=E_{cm}~\sigma(E_{cm})~e^{2\pi \eta}
\end{equation}
Where $\sigma$(E$_{cm}$) is the measured cross-section at centre of mass energy, E$_{cm}$. The second term ($e^{2\pi \eta}$) removes the strong energy dependence of $\sigma$. It accounts for the s$-$wave Coulomb barrier transmission at low energies with $\eta$ being the Sommerfeld parameter, defined as,
\begin{equation}
\eta=\frac{z_1z_2e^2}{\hbar} \sqrt{\frac{\mu}{2E}}
\end{equation}
z$_1$, z$_2$ being the charges of the target and projectile and $\mu$ is the reduced mass. S-factor calculated from measured cross-section data have been listed in Table~\ref{sfactor_list} and shown in Figure~\ref{sfactor_fig}.
\begin{table}
\centering
\caption{\label{sfactor_list}Calculated S-factors from measured cross-sections.}
\footnotesize
\begin{tabular}{@{}cc}
\br
Energy (E$_{cm}$) in MeV&S-factor ($\times 10^{10}$) in MeV-b\\
\br
6.73	$\pm$	0.02	&	0.198	$\pm$	0.050	\\
6.42	$\pm$	0.03	&	0.206	$\pm$	0.051	\\
5.84	$\pm$	0.05	&	0.227	$\pm$	0.060	\\
5.48	$\pm$	0.05	&	0.295	$\pm$	0.074	\\
5.09	$\pm$	0.06	&	0.318	$\pm$	0.080	\\
4.66	$\pm$	0.07	&	0.567	$\pm$	0.167	\\
4.23	$\pm$	0.08	&	0.728	$\pm$	0.262	\\
4.08	$\pm$	0.09	&	0.816	$\pm$	0.368	\\
3.65	$\pm$	0.10	&	0.952	$\pm$	0.429	\\
3.15	$\pm$	0.11	&	1.020	$\pm$	0.464	\\
2.57	$\pm$	0.13	&	2.542	$\pm$	1.152	\\

\br
\end{tabular}\\
\end{table} 

\subsection{Thermonuclear reaction rate prediction}
\begin{figure}[h]
\centering
\includegraphics[clip, trim=0.0cm 0.0cm 0.0cm 0.0cm,width=0.7\textwidth]{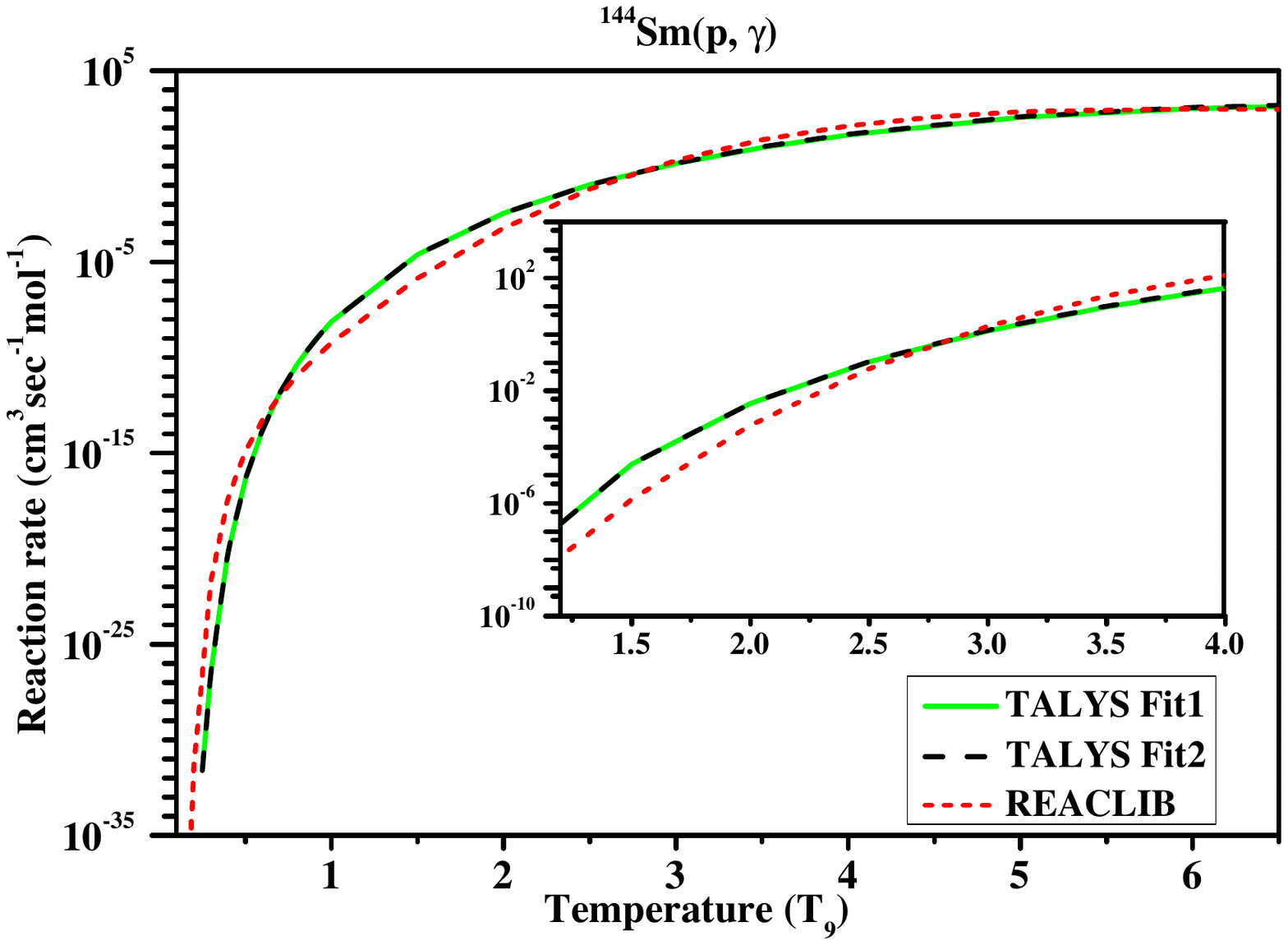}
\caption{Reaction rate estimated for $^{144}$Sm(p, $\gamma$) reaction from fitting parameters obtained from TALYS. Reaction rate value given in the REACLIB database has been shown for the comparison purpose.}
\label{144Sm_prate}
\end{figure}

\begin{figure}[h]
\centering
\includegraphics[clip, trim=0.0cm 0.0cm 0.0cm 0.0cm,width=0.7\textwidth]{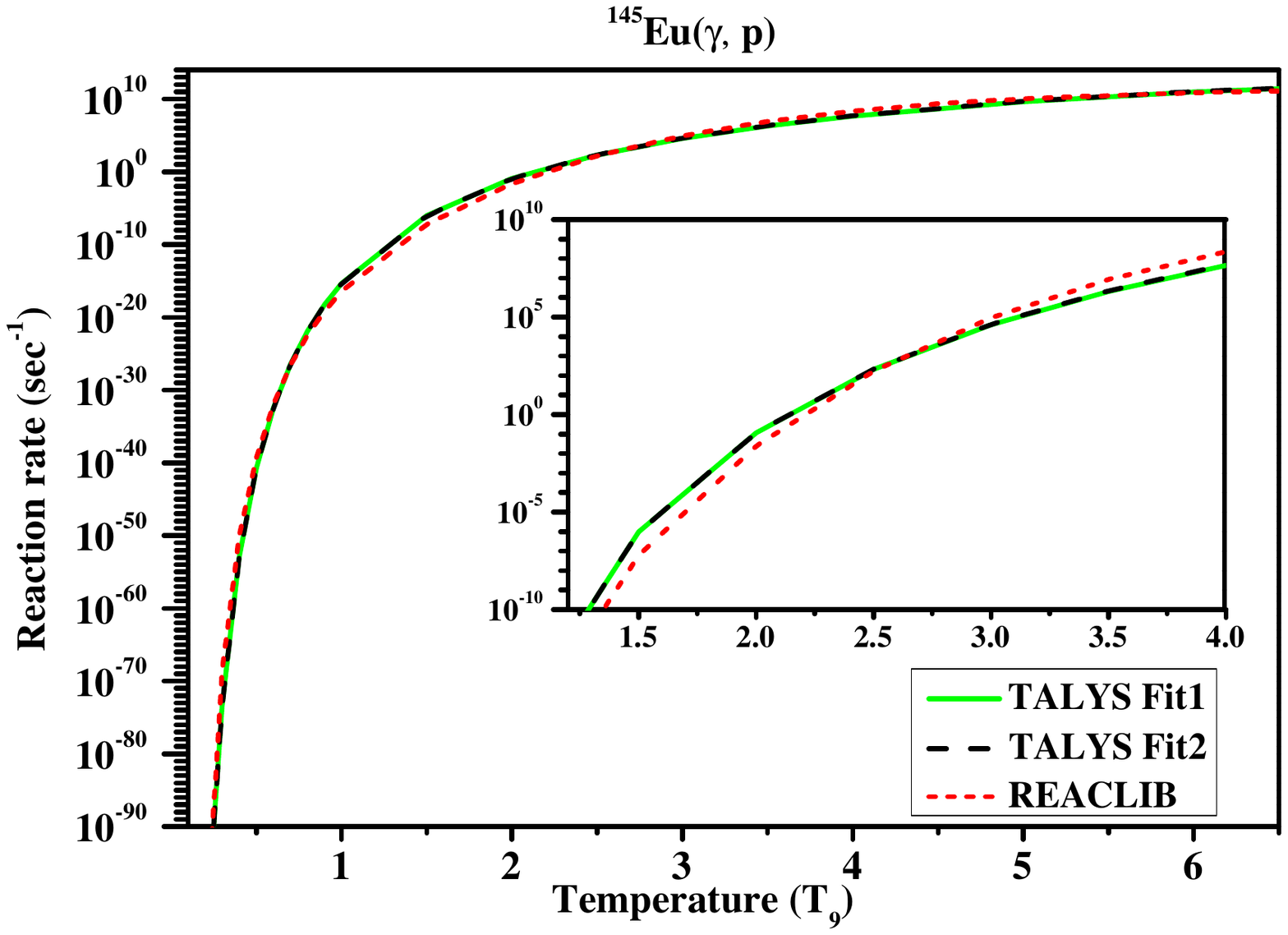}
\caption{Reaction rate estimated for $^{145}$Eu($\gamma$, p) reaction from $^{144}$Sm(p, $\gamma$) reaction rate and reciprocity theorem (TALYS+RT). Reaction rate value given in the REACLIB database has been shown for the comparison purpose.}
\label{rate145eu_gp}
\end{figure}

The reaction rates (R) for the ground state $^{144}$Sm were estimated from the best fitted parameters (TALYS Fit1 and TALYS Fit2) of the $^{144}$Sm(p, $\gamma$) cross-section using TALYS 1.96. These values were compared with the REACLIB database~\cite{reaclib}, shown in Figure~\ref{144Sm_prate}.
Reaction rates obtained for $^{144}$Sm(p, $\gamma$) $^{145}$Eu reaction can be used to find the rate for the inverse process $^{145}$Eu($\gamma$, p), by applying $``$reciprocity theorem"~\cite{rolfs1988cauldrons, holmes1976tables, arnould2003p}.
A compact form of the theorem for a reaction I(j, $\gamma$)L can be written as Eq.~\ref{rate_eq},
\begin{equation}
R_{\gamma j}=9.8677\times10^9~T_9^{3/2} (\frac{g_I g_j}{g_L}) (\frac{G_I}{G_L}) (\frac{A_I A_j}{A_L})^{3/2} N_A \bigl \langle\sigma v\bigr \rangle_{j\gamma}~e^{-\frac{11.605~Q_{j\gamma}(MeV)}{T_9}}
\label{rate_eq}
\end{equation}
where g$_n=$2J$_n^0$+1, J$^0$ being the ground state spin, G is the temperature dependent normalised partition function, N$_A$ and A denotes Avogadro number and mass of the nuclei, Q$_{j\gamma}$ is the Q-value of the (j, $\gamma$) reaction and $\bigl \langle\sigma v\bigr \rangle_{j\gamma}$ represents reaction rate per particle pair. 
Estimated reactions rates obtained using TALYS and reciprocity theorem for ($\gamma$, p)$^{144}$Sm are shown in Figure~\ref{rate145eu_gp}.

\section{Conclusion}
The cross-section measurement of the proton capture reaction with $^{144}$Sm target at energies between 2.6 and 6.8 MeV was performed. The experiment was carried out at VECC K130 Cyclotron facility, Kolkata using a stack foil activation technique. Astrophysically relevant energy regions corresponding to temperature 3, 4 GK and a partial portion of 2 GK were explored and $S-$factors were measured for all energies and for the first time at E$^{cm}_p=$2.57$\pm$0.13 MeV ($S-$ factor$=$2.542(1.152)$\times$10$^{10}$ MeV-b). Molecular deposition method has been used to prepare the enriched $^{144}$Sm targets. Measured cross-section data was compared with the statistical model code TALYS 1.96. Best fitted parameters like OPM, NLD and $\gamma-$SF were used to calculate the reaction rates of $^{144}$Sm(p, $\gamma$) reaction using TALYS. Further these reaction rate values were utilised to estimate the reaction rate of inverse process, $^{145}$Eu($\gamma$, p) using TALYS and reciprocity theorem.
\section*{Acknowledgement}
The authors are extremely thankful to Prof. Chandi Charam Dey for providing enriched target material and Prof. Chandana Bhattacharya towards the successful execution of the experiment. A sincere thanks to Mr. Sudipta Barman and other workshop members of Saha Institute of Nuclear Physics, Kolkata for their support. We acknowledge the kind support provided by the staffs of K130 Cyclotron at VECC, Kolkata and Mr. A. A. Mallick of Analytical Chemistry Division, BARC; VECC for his assistance during irradiation experiment. SS would like to thank the Council of Scientific and Industrial Research (CSIR), Government of India, for financial support in the form of Senior Research Fellowships (File No 09/489(0119)/2019-EMR-I).  
\section*{Data availability statement}
The data that support the findings of this study are available upon reasonable request from the authors.
\section*{References}

\bibliography{ref_file}

\end{document}